\documentclass[twocolumn]{revtex4}
\usepackage{amsmath,graphicx}

\begin{document}
\title{Invalidation of the Kelvin Force in Ferrofluids}
\author{Stefan Odenbach$^{1}$ and Mario Liu$^{2}$~\cite{email}
\\ $^1$ZARM, Universit\"{a}t Bremen, 28359 Bremen, Germany\\
$^2$Theoretische Physik, Universit\"{a}t Hannover, 30167
Hannover, Germany}
\date{\today}
\begin{abstract}
Direct and unambiguous experimental evidence for the
magnetic force density being of the form
$M{\boldsymbol\nabla} B$ in a certain geometry --- rather
than being the Kelvin force $M{\boldsymbol\nabla} H$ ---
is provided for the first time. ($M$ is the
magnetization, $H$ the field, and $B$ the flux density.)
\end{abstract}
 \maketitle

In studying polarizable and magnetizable condensed
systems, one of a few key concepts is the force exerted
by applied electric and magnetic fields. A clear and
precise understanding here is therefore of crucial
importance. For a neutral, dielectric body, this force is
usually taken to be the Kelvin force~\cite{LL,si,n1},
\begin{equation}\label{kelvin}
{\pmb {\cal F}}_{\rm K}=\int [P_i{\boldsymbol\nabla}E_i+
\mu_0M_i{\boldsymbol\nabla}H_i]\,{\rm d}^3r,
\end{equation}
where $P$ is the polarization, $M$ the magnetization, $E$
the electric and $H$ the magnetic field; $\mu_0$ and
$\epsilon_0$ are as usual the vacuum permeability and
permittivity, and the integral is over the volume of the
body. (Summation over repeated indices is always
implied.)

In spite of general believes, however, the theoretical
foundation for this expression is far from rock solid,
and recent considerations have pointed to the validity,
under conditions to be specified below, of the variant
expression~\cite{rev}
\begin{equation}\label{kelvin2}
{\pmb {\cal F}}_{\rm v}=\int
[(P_i/\epsilon_0){\boldsymbol\nabla}D_i+
M_i{\boldsymbol\nabla}B_i]\,{\rm d}^3r,
\end{equation}
where $\bf D\equiv\epsilon_0E+P$ and $\bf B\equiv\mu_0
(H+M)$ are the dielectric displacement and magnetic flux
density, respectively. This circumstance makes a direct
measurement of the electromagnetic force desirable.

\begin{figure}
\begin{center}
\includegraphics[width=0.63\columnwidth]{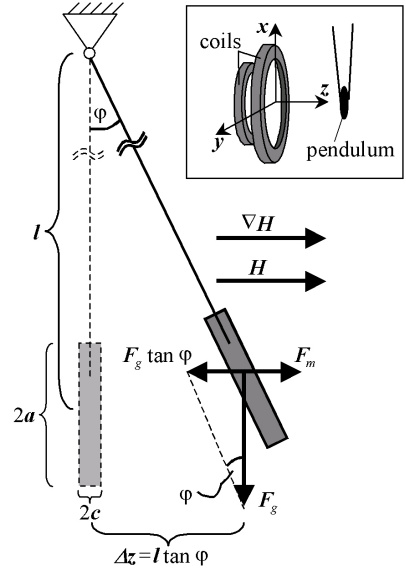}
\end{center}
\caption{Schematic representation of the experiment, a
pendulum deflected by a magnetic force. The inset shows
the two magnetic coils producing the field. The object
marked as pendulum is a disk containing ferrofluid, with
its face perpendicular to both the field and field
gradient.}\label{Fig1}
\end{figure}

In this Letter, such an experiment for the magnetic part
of the force is reported. It consists of a pendulum
exposed to a well defined, horizontal magnetic field with
a gradient parallel to the field, see Fig. 1. The weight
of the pendulum is a disk-shaped container. It is filled
with ferrofluid and has its face aligned normal to the
field. The disk is drawn by the field gradient, but
leaves the applied field essentially unchanged. Given the
careful design of the experiment, with especially the
string of the pendulum close to 8 m, its displacement
yields an easy and accurate measurement of the magnetic
force. For the given geometry,  Eq~(\ref{kelvin2}) was
verified to great precision.

The quest for the correct form of the electromagnetic
force sounds, and is indeed, vintage 19$^{\rm th}$
century. So it is truly amazing that something as
elementary as the rest position of a pendulum subject to
a static magnetic field cannot be predicted in general
consensus even today. Kelvin was the first to have made a
lasting contribution in this context by first
calculating, microscopically, the force exerted by an
external field on a single dipole, and then taking the
total force as the sum of the forces on all dipoles.
Clearly, this is only justified if the dipoles are too
far apart to interact, if the system is
electromagnetically dilute. This implies a small
magnetization and susceptibility, $M\ll H$, $\chi\ll1$.
(We shall only be discussing the magnetic force from here
on.) Part of the terms neglected must be quadratic in
$M$, as they naturally account for interaction.
Therefore, this approach cannot distinguish between
Eq~(\ref{kelvin}) and (\ref{kelvin2}), the difference of
which is $\frac{1}{2}\mu_0 {\boldsymbol\nabla}M^2$, a
term quadratic in $M$, or $\chi$.

The classic derivation of the electromagnetic force is
thermodynamic in nature, and rather more generally valid,
cf \S15 and 35 of the book on macroscopic electrodynamics
by Landau and Lifshitz ~\cite{LL}. Starting from the
Maxwell stress tensor $\Pi_{ij}$, this derivation
subtracts the hydrostatic ``zero-field pressure" -- the
pressure that would be there in the absence of fields,
for given temperature and density -- and identifies the
gradient of the rest as the electromagnetic (or Helmholz)
force. If the susceptibility $\chi$ -- irrespective of
its value -- is taken to be proportional to the density
(or concentration of magnetic particles in a suspension
such as ferrofluid), this expression reduces to the
Kelvin force, Eq~(\ref{kelvin}). With the difference
between Eq~(\ref{kelvin}) and (\ref{kelvin2}) appreciable
if $\chi\gtrsim1$, this derivation may be taken to
clearly rule out Eq~(\ref{kelvin2}).

Ferrofluids are stable suspensions of magnetic
nano-particles, with an initial susceptibilities up to
about 5 that are usually proportional to the volume
concentration $\rho_1$ of magnetic particles at small
fields, see Fig 2. It is therefore standard assumption
that the Kelvin force, Eq~(\ref{kelvin}), may be employed
wherever $\chi\sim\rho_1$ holds~\cite{rs}.
\begin{figure}
\begin{center}
\includegraphics[width=0.95\columnwidth]{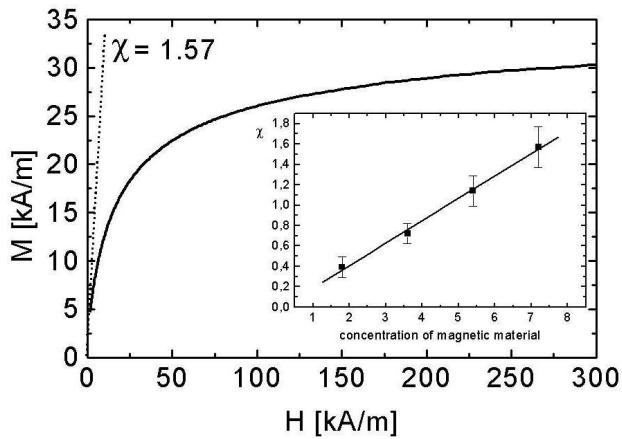}
\end{center}
\caption{Magnetization of a typical ferrofluid (No 7 in
Table 2) as a function of the field $H$. The inset shows
the dependence of $\chi$ on the concentration of magnetic
particles for four fluids of similar microscopic make-up,
especially the mean particle diameter.}\label{Fig2}
\end{figure}

Closer examination~\cite{luo}, however, has raised doubts
about the validity of Eq~(\ref{kelvin}) for highly
magnetizable systems, by unearthing an implicit
assumption of small susceptibilities, made during the
derivation as given by Landau and Lifshitz. So, in spite
of appearance, neither does this derivation distinguish
between Eq~(\ref{kelvin}) and (\ref{kelvin2}). Rather
more important, however, is the fact that this force is
not operative in equilibrium, when the divergence of the
Maxwell stress vanishes, $\nabla_j\Pi_{ij}=0$, and the
gradient of the zero field pressure compensates the
Kelvin force.

Generally speaking, while the electromagnetic force is a
simple and unique quantity in microscopic physics, given
by the Lorentz force, it is a multi-faceted, heuristic
concept in macroscopic physics. Not aware of this fact,
and possibly for want of something better, we tend to
take Eq~(\ref{kelvin}) -- with its curious preference of
$E, H$ over $D, B$ -- as {\em the} electromagnetic force.
(There is of course also the macroscopic Lorentz force,
which is however zero if the system is neutral and
insulating).

To understand why this is a conceptual pitfall, consider
the hydrodynamic theory for ordinary fluids, such as
water or air, in the absence of fields. It consists of a
set of partial differential equations including the
Navier-Stokes equation~\cite{LL6}. By solving these
equations with the appropriate initial and boundary
conditions, one is in principle able to predict any
experimental outcome -- say the trajectory of an airplane
-- without ever the need to introduce the concept of
force. For the convenience of reasoning and arguing,
however, we do label certain terms as force densities: in
the differential equations, boundary conditions, or the
solution under consideration. As a result, these forces
depend on the context and geometry -- take the form of
the airfoil and the associated lift.

Circumstances are similar in the hydrodynamic theory of
ponderable systems~\cite{LL,rev,ml}, and there is just as
little reason to expect the existence of a unique
electromagnetic force that is independent from geometry
and context. But we do also need the concept of force:
Physics does not comprise solely of mathematics, and
clear thinking about the valid expressions for the
electromagnetic force under given conditions will give us
considerable heuristic power in prediction --- without
having to solve the set of partial differential equations
each and every time. A simple question about the rest
position of a pendulum subject to a field should be
answerable by considering an elementary force
equilibrium, between the strain in the string, the
gravitational and the electromagnetic force. Solving the
hydrodynamic theory with appropriate boundary
conditions~\cite{rev} and assuming linear constitutive
relation, the magnetic force is found to be
\begin{equation}\label{f}
{\pmb {\cal F}}={\textstyle\frac{1}{2}}\oint[\mu_0\chi
H{\rm _t}^2 +\chi B{\rm _n}^2/(1+\chi)\mu_0]\,{\rm d}{\bf
A},
\end{equation}
where the surface integral is to be taken over the
surface of the magnetizable body. The subscripts $_{\rm
t}$ and $_{\rm n}$ denote the tangential and normal field
component with respect to the surface. As the above
fields are continuous across the interface, they may be
both the external or the internal field. Eq~(\ref{f}) is
rather generally valid. It holds irrespective of $\chi$'s
size, or its dependency on either the density or the
volume concentration of magnetic particles; it is also
easily generalized for nonlinear constitutive relations.
Equilibrium, however, is an essential requirement.

Employing the Gauss law, and provided $\chi$ is spatially
constant, with $M=\chi H=$ $\chi B/\mu_0(1+\chi)$, we may
rewrite Eq~(\ref{f}) as a volume integral,
\begin{equation}\label{ff}
{\pmb {\cal F}} = \int(\mu_0M_{\rm t}
{\boldsymbol\nabla}H_{\rm t} +M{\rm _n}
{\boldsymbol\nabla}B_{\rm n})\,{\rm d}^3r.
\end{equation}
If the field is normal to the surface, as in the above
experiment, the first term vanishes, and ${\pmb {\cal
F}}_{\rm v}$ of Eq~(\ref{kelvin2}) is retrieved; if the
field is tangential to the surface, Eq~(\ref{kelvin}) and
${\pmb {\cal F}}_{\rm K}$ hold.

Casting doubts -- however well theoretically founded --
on the applicability of the Kelvin force certainly
requires experimental validation. Fortunately, in
strongly magnetizable systems such as ferrofluids, the
difference between ${\pmb {\cal F}}_{\rm K}$ and ${\pmb
{\cal F}}_{\rm v}$ is appreciable. For linear
constitutive relations, the latter is larger by the
factor $1+\chi$: With the magnetic field perpendicular to
the disk surface, we have $B_{\rm in}=B_{\rm ex}$
$=\mu_0H_{\rm ex}$, where the subscripts $_{\rm in}$ and
$_{\rm ex}$ denote the internal and external field,
respectively. This implies $H_{\rm in}=$ $H_{\rm
ex}/(1+\chi)$ and $M=\chi H_{\rm in}$ $=\chi H_{\rm
ex}/(1+\chi)$. As the disk containing the ferrofluid is
not infinite, the external field $H_{\rm ex}$ is not
quite the applied field $H_0$. Approximating the disk as
a (flat) ellipsoid, we may write $H_{\rm ex}/(1+\chi)$
$=H_{\rm in}=$ $H_0/(1+D\chi)$, where $D$ is the
demagnetization factor. (The ferrofluid sample has the
form of a flat disk with a diameter of $a = 50$~mm,
thickness $c = 1$~mm, and $b\equiv a/c$ $=50$, from which
the demagnetization factor in $z$ direction is calculated
as $D=0.9694$, employing the approximation~\cite{5}
$D=1-$ $\pi/2b$ $+2b^2$.) Inserting these formulas into
Eqs~(\ref{kelvin}) and (\ref{kelvin2}), respectively, we
find
\begin{eqnarray}\label{f2}
{\pmb{\cal F}}_{\rm
K}=[{\textstyle\frac{1}{2}}V\mu_0\chi/(1+D\chi)^2]
{\boldsymbol\nabla}H_0^2, \\ \label{f1} {\pmb{\cal
F}}_{\rm v}=(1+\chi) [{\textstyle\frac{1}{2}}V\mu_0\chi
/(1+D\chi)^2] {\boldsymbol\nabla}H_0^2.
\end{eqnarray}
They show ${\pmb{\cal F}}_{\rm v}=(1+\chi) {\pmb{\cal
F}}_{\rm K}$, and give the force in terms of simple,
measurable quantities: susceptibility $\chi$, applied
field $H_0$, the demagnetization factor $D$, and the
ferrofluid volume $V$. (Due to the small gradient of the
applied field, the volume integral was approximated by
multiplication with $V$.)

\begin{center}
\begin{tabular}{|c|c|c|}\hline
description &symbol&value\\\hline
 length of pendulum &l &
7650mm\\
 diameter of fluid disk&a&50mm\\
 thickness of fluid disk&c&1mm\\
 volume of fluid disk&V&2ml\\
mass of pendulum&m&35g\\
 demagnetization
factor&D&0.9694\\ field parameters&$\alpha_1$&$2.5\times
10^4{\rm m}^{-3}$\\
   &$\alpha_2$&$0.47$m\\
   &$\alpha_3$&$750{\rm m}^{-1}$\\ \hline
\end{tabular}

\vspace{3mm}
 Table 1. Technical data of the experiment
\end{center}

The magnetic field, with its direction and gradient
both parallel to the $z$-direction, is produced by an
arrangement of two coils, which is simply half a
Fanselau-arrangement~\cite{6}, see Fig. 1. The field
is homogeneous in the $xy$-plane to an accuracy of
better than 0.1 \%. Both the field and its gradient
are proportional to the applied current $I$. The
field strength shows a quadratic decrease in
$z$-direction, leading to a linear dependence of the
field gradient. Choosing the center of the larger
coil as the origin of the coordinate system ($z=0$ is
at the outer edge of the coil), the field is
parameterized with three coefficients,
$\alpha_1=2.5\times10^4{\rm m}^{-3}$,
$\alpha_2=0.47$m, $\alpha_3=750{\rm m}^{-1}$, as
\begin{equation}\label{H0}
H_0=[\alpha_1(z-\alpha_2)^2-\alpha_3]I.
\end{equation}
($H_0$ denotes the $z$-component of the field. The other
components are not zero for $z\not=0$, as
${\boldsymbol\nabla}\cdot{\bf B}=0$. The force density
from the $x,y$-components of the field is quadratically
small, and integrates to zero for a disk centered at
$z=0$.)

The disk containing the ferrofluid is attached to two
nylon strings, of length $l=7650$~mm and mounted to a
tripod, forming a pendulum. The rest position of the
pendulum is at $x,y=0$ and $z=7.6$~cm. If a magnetic
field is applied, the magnetic force acting on the fluid
attracts the disk towards the coils. This results in its
displacement and, as shown in Fig.~1, a restoring
gravitational force, the relevant component of which is
$mg\tan\varphi$ $=mg\Delta z/l$, with $m$ the mass of the
disk, $g$ the gravitational acceleration, $\varphi$ the
angle of deflection, and $\Delta z$ the displacement. In
equilibrium, the gravitational force is equal to the
magnetic one, either ${\pmb{\cal F}}_{\rm K}$ of
Eq~(\ref{f2}) or ${\pmb{\cal F}}_{\rm v}$ of
Eq~(\ref{f1}), but always $\sim\nabla H_0^2$. Writing the
magnetic force as $\beta \nabla H_0^2$ and equating it
with $mg\Delta z/l$, we find
\begin{equation}\label{exp}
\Delta z=(\beta l/mg)\nabla H_0^2= \lambda H_0\nabla H_0.
\end{equation}
The slope $\lambda\equiv2\beta l/mg$ is measured by
plotting $\Delta z$ over $H_0\nabla H_0$.

Due to the great length of the pendulum, even small
changes of the magnetic force will provide easily
measurable changes of $\Delta z$. For instance, a
change of the magnetic force of $10^{-5}$N
corresponds to $\Delta z=1$mm. In addition, since
$\Delta z=13$ mm corresponds to $\varphi=0.1^\circ$,
even relatively large displacements, of the order of
a few cm, will not lead to a tilt of the disk
relative to the field direction. This ensures the
homogeneity, over the sample, of the magnetic field
$H_0$ in x-direction. The position of the sample is
monitored by a digital video system, allowing a
determination of the displacement $\Delta z$ with an
accuracy of approximately 0.2~mm.

\begin{center}
\begin{tabular}{|c|c|c|c|c|}
  \hline
  ferrofluid & $\chi$ & concentration & particle size&
  trade name\\ \hline
  1 & 0.183 & 1.8 vol\%& 7.1 nm& APGS20n\\
  2 & 0.393 & 1.8 vol\%& 9.2 nm& APG510A\\
  3 & 0.543 & 1.4 vol\%& 9.2 nm&  \\
  4 & 0.723& 3.7 vol\%& 9.2 nm&APG511A \\
  5 & 1.141& 5.4 vol\%& 9.1 nm&APG512A \\
  6 & 1.27 & 6.2 vol\%& 8.7 nm&   \\
  7 & 1.57 & 7.3 vol\%& 9.2 nm&APG513A\\
  8 & 1.84 & 7.3 vol\%& 8.9 nm&  \\
  9 & 2.27& 7.3 vol\%& 10.4 nm&EMG905 \\
  10 & 3.65 & 13 vol\%& 8.5 nm&  \\
  11 & 3.95 & 16.4 vol\%& 9.5 nm&EMG900 \\ \hline
\end{tabular}

\vspace{3mm}
 Table 2. Data of the employed ferrofluids
\end{center}

With the experimental setup described above, we have
carried out a series of experiments using 11 different
ferrofluids having different initial susceptibility,
ranging from 0.18 to approximately 4, cf table~2. (For
ferrofluids supplied by Ferrofluidics, the trade names
are given in table 2.) Because only weak magnetic fields
up to a maximum of $H_0=10$kA/m are used in the
experiment, the validity of linear constitutive relation
is ensured, cf Fig 2, and we can use the initial
susceptibility of table 2 throughout the evaluation of
the data.

\begin{figure}
\begin{center}
\includegraphics[width=0.95\columnwidth]{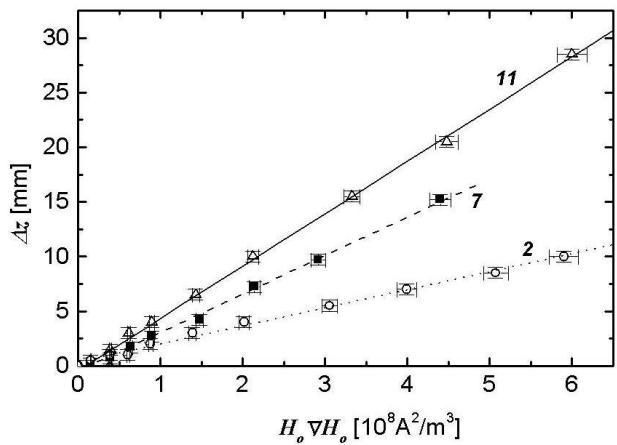}
\end{center}
\caption{Dependence of the displacement $\Delta z$ of the
pendulum on $H_0\nabla H_0$, for fluids 2, 7 \& 11, with
linear fits.}\label{Fig3}
\end{figure}

Fig.~3 shows the displacement $\Delta z$ of the disk as
functions of $H_0\nabla H_0$, for three different fluids
from table 2. It is obvious that the linear relation
between $\Delta z$ and $H_0\nabla H_0$ is well satisfied
over the whole range of investigation, attesting to the
validity of the linear constitutive relation employed in
the evaluation. From plots like these, we can determine
the slope to an accuracy of about 3\%. By averaging over
five runs for each fluid, the accuracy is increased to
approximately 1.5\%, demonstrating the reproducibility of
the data.

\begin{figure}
\begin{center}
\includegraphics[width=0.95\columnwidth]{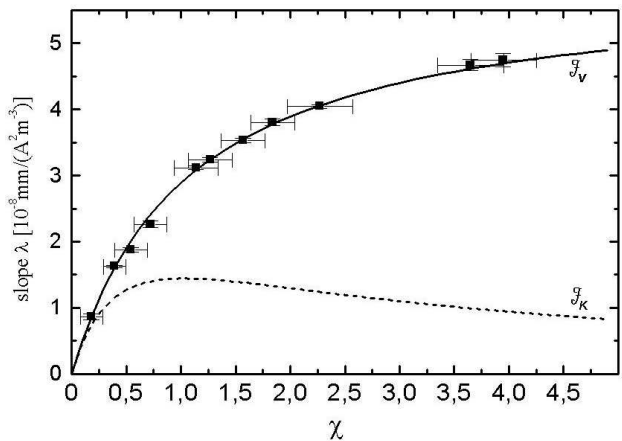}
\end{center}
\caption{$\lambda=\Delta z/(H_0\nabla H_0)$ as a function
of the susceptibility. The dashed line is calculated
employing the Kelvin force ${\pmb{\cal F}}_{\rm K}$, of
Eq~(\ref{f2}), while the solid line is calculated with
the variant form ${\pmb{\cal F}}_{\rm v}$, of
Eq~(\ref{f1}).}\label{Fig4}
\end{figure}

Contact with theory is made by plotting $\lambda$ as a
function of $\chi$ for all 11 fluids described in
table~2, see Fig.~4. The solid line gives $\lambda$ as
calculated with ${\pmb{\cal F}}_{\rm v}$ of
Eq~(\ref{f1}), while the dashed line is calculated with
${\pmb{\cal F}}_{\rm K}$ of Eq~(\ref{f2}) -- both using
the data of table~1 and the independently measured
susceptibility of table~2.

Clearly, the agreement between the solid line and the
experimental data is excellent. Moreover, it is obvious
that the error committed by using the Kelvin force
indiscriminately can be considerable. As the correction
is $1+\chi$, however, this is an issue only for highly
magnetizable systems, say with susceptibilities
$\gtrsim0.2$, which helps to explain why this was not
seen earlier.

\vspace{25mm}
 \noindent{\bf Acknowledgment:} We would
like to thank Hanns-Walter M\"{u}ller for a critical reading
of the manuscript and for helpful comments.


\begin{thebibliography}{99}
\bibitem{n1} The Kelvin foce is sometimes given as
$P_i\nabla_i{\bf E}+$ $M_i\nabla_i{\bf H}$, which differ
from Eq~(\ref{kelvin}) by terms
$\sim{\boldsymbol\nabla}\times{\bf E}$,
${\boldsymbol\nabla}\times{\bf H}$, both zero if the
applied field is static, and the medium insulating.
\end{thebibliography}
\end{document}